\documentclass[10pt,journal,compsoc]{IEEEtran}

\ifCLASSOPTIONcompsoc
  \usepackage[nocompress]{cite}
\else
  \usepackage{cite}
\fi

\usepackage{booktabs} 
\usepackage[ruled]{algorithm2e} 
\usepackage[T1]{fontenc}
\usepackage{amsmath}
\usepackage{multirow}
\usepackage{array}
\usepackage{color}
\usepackage{listings}
\usepackage{fancyvrb}
\usepackage{cuted}
\usepackage{inputenc}
\usepackage{placeins}
\usepackage{amsmath, amssymb, tabularx}
\usepackage{lipsum}
\usepackage{hyperref}
\usepackage{tabto}
\usepackage{setspace}
\usepackage{balance}
\usepackage{enumitem}
\usepackage{longtable}
\usepackage{pdflscape}
\usepackage{pdflscape}
\usepackage{wrapfig}
\usepackage{rotating}
\usepackage{makecell}
\usepackage{array}
\usepackage{pdflscape}
\usepackage{afterpage}
\usepackage{svg}
\usepackage{color}
\usepackage{colortbl}
\usepackage{graphicx}
\usepackage{balance}

\begin{document}

\title{ScissionLite: Accelerating Distributed Deep Neural Networks Using Transfer Layer}

\author{Hyunho~Ahn,
        Munkyu~Lee,
        Cheol-Ho~Hong,
        and~Blesson~Varghese
\IEEEcompsocitemizethanks{\IEEEcompsocthanksitem Hyunho Ahn, Munkyu~Lee, and Cheol-Ho~Hong (corresponding author) are with the School of Electrical and Electronics Engineering, Chung-Ang University, Seoul, Korea. e-mail: cheolhohong@cau.ac.kr
\IEEEcompsocthanksitem Blesson Varghese is with with the School of Electronics, Electrical Engineering and Computer Science, Queen's University Belfast, United Kingdom.}
}

\markboth{05 May 2021}
{Ahn \MakeLowercase{\textit{et al.}}: ScissionLite: Accelerating Distributed Deep Neural Networks Using Transfer Layer}

\IEEEtitleabstractindextext{%
\begin{abstract}
Industrial Internet of Things (IIoT) applications can benefit from leveraging edge computing. For example, applications underpinned by deep neural networks (DNN) models can be sliced and distributed across the IIoT device and the edge of the network for improving the overall performance of inference and for enhancing privacy of the input data, such as industrial product images. However, low network performance between IIoT devices and the edge is often a bottleneck. In this study, we develop ScissionLite, a holistic framework for accelerating distributed DNN inference using the Transfer Layer (TL). The TL is a traffic-aware layer inserted between the optimal slicing point of a DNN model slice in order to decrease the outbound network traffic without a significant accuracy drop. For the TL, we implement a new lightweight down/upsampling network for performance-limited IIoT devices. In ScissionLite, we develop ScissionTL, the Preprocessor, and the Offloader for end-to-end activities for deploying DNN slices with the TL. They decide the optimal slicing point of the DNN, prepare pre-trained DNN slices including the TL, and execute the DNN slices on an IIoT device and the edge. Employing the TL for the sliced DNN models has a negligible overhead. ScissionLite improves the inference latency by up to 16 and 2.8 times when compared to execution on the local device and an existing state-of-the-art model slicing approach respectively. 
\end{abstract}

}

\maketitle

\IEEEdisplaynontitleabstractindextext

\IEEEpeerreviewmaketitle

\IEEEraisesectionheading{\section{Introduction}\label{sec:introduction}}
\IEEEPARstart{E}{dge} computing is a comparatively new computing paradigm that leverages computation resources at the edge of a network. A local device that offloads its workload to the edge is then able to achieve lower latency than cloud computing~\cite{hong2019resource, hong2018qcon, wang2017enorm}. Edge computing is currently gaining growing interest in Industrial Internet of Things (IIoT) applications. IIoT applications usually employ different types of industrial resources through the network to construct a service-oriented industrial ecosystem~\cite{qiu2020edge}. Edge computing in IIoT allows data generated from industrial machines, devices, and sensors to be processed with low latency and thus enables time-critical decision-making.

Recent advances in edge computing for IIoT are best exemplified by deep neural network (DNN) model slicing for product surface inspection~\cite{zeng2019boomerang}. This research employs a surface inspection camera on an automated assembly line and captures the surface of products one by one for DNN inference. However, as IIoT devices have limited computation capability, the inspection time can be continuously delayed, resulting in a significant decline of outcomes. To address this issue, the research splits the entire DNN model and distributes each slice into the device and the edge. Leveraging both device and edge resources for DNN inference has the benefit of preserving privacy for input data and also reducing inference time.

\emph{Scission}~\cite{lockhart2020scission}, our previous work, generates a list of slicing points of a DNN model against different user-defined objectives for minimizing inference time. Scission automatically benchmarks the DNN model on the device, edge, and cloud resource. It then analyzes the benchmark data including the execution time of respective DNN layers and the data transfer time. Scission decides the optimal point for model slicing based on the following criteria: (1) Computationally intensive layers should be placed on a capable resource, and (2) minimum data transfer is required between the device, edge, and cloud. Scission discovers the intersection point of the two policies that is best suited for the lowest inference time.

Scission can achieve optimal performance for distributed deep learning inference. However, the overhead of data transfer between the local and the edge is still problematic in IIoT applications. IIoT now endeavors to exploit multi-access edge computing (MEC) that enables cloud computing at the edge of the cellular network~\cite{qiu2020edge, hou2019iiot, sun2018double}. 5G cellular networks provide high download bandwidth ranging from 600 to 1,700 Mbps~\cite{narayanan2020first}. However, the upload speed is comparatively low (30 -- 60 Mbps) owing to weak cellular radio power on devices. Therefore, when the data of an IIoT device is uploaded to the multi-access edge, the limited network bandwidth can be a significant bottleneck.

Although existing research presents methods to reduce the data transfer amount between the device and the edge by applying a  data encoding method to the split point of the DNN~\cite{eshratifar2019bottlenet, hu2020fast, shao2020bottlenet++, matsubara2021split}, they are limited in the following ways: (1) The methods are usually optimized for a very slow network environment (e.g., 50 Kbps to 20 Mbps)~\cite{eshratifar2019bottlenet, matsubara2021split}. Therefore, it is not clear whether the proposed methods would suit faster 5G networks. (2) The studies adopt compression techniques that are not integrated into the original DNN~\cite{eshratifar2019bottlenet} or that incur high computational complexity~\cite{hu2020fast, shao2020bottlenet++}. In these cases, the compression techniques would be a further bottleneck for performance-limited IIoT devices. (3) The methods do not find the optimal slicing point when the traffic is reduced since they are based on estimations rather than empirical data. (4) The studies are validated on lightweight DNN models such as MobileNetV2 and VGG16 with relatively few layers and small datasets~\cite{eshratifar2019bottlenet, hu2020fast, shao2020bottlenet++} such as CIFAR\footnote{\url{https://www.cs.toronto.edu/~kriz/cifar.html}}. Small models and input images may not be suitable for IIoT applications that use high-quality images as input and demand high inference accuracy. 

In this paper, we propose \emph{ScissionLite}, a framework to accelerate edge-based distributed deep learning inference for IIoT. ScissionLite inserts an additional traffic-aware layer called the Transfer Layer (TL) between the optimal slicing point of the DNN model in order to decrease the amount of data transferred between the device and the edge without significantly sacrificing the accuracy of inference. The TL embeds a small neural network between the split point that compresses the output data of the IIoT device and expands the compressed data on the edge. For this purpose, we develop a new lightweight TL using a down/upsampling network that is integrated into the original DNN for performance-limited IIoT devices. 

ScissionLite is a holistic framework for edge-based distributed inference that consists of ScissionTL, the Preprocessor, and the Offloader. ScissionTL is a Scission-based tool for automated benchmarking that can decide the optimal slicing point when the traffic is reduced by the TL. The Preprocessor creates and retrains a new DNN model that employs the TL. Finally, the Offloader is a runtime platform that distributes deep learning inference into the IIoT device and the edge. For evaluation, we utilize five large-scale pre-trained production DNNs, namely DenseNet169, DenseNet201, ResNet101, InceptionResNetV2, and InceptionV3\footnote{\url{https://keras.io/api/applications/}}, with the ImageNet dataset\footnote{\url{https://image-net.org}} in an emulated 5G environment. It is noted that using TL introduces a negligible overhead. ScissionLite can increase the performance of deep learning inference by up to 16 and 2.8 times compared to the local device execution and original Scission respectively with 0.9 -- 1.4\% of an accuracy drop.

The contributions of this article are summarized as follows:

\begin{itemize}
    
    \item We design ScissionLite and perform its evaluation using large-scale production DNNs with a high-precision data set in an emulated 5G environment. To the best of our knowledge, this is the first work that shows the feasibility of transfer layer-based distributed DNNs in a realistic edge-based IIoT environment.
    
    \item We implement a new lightweight TL using a down/upsampling neural network suitable for low power processors in IIoT devices. This simple network is seamlessly integrated into the target DNN, so that it neither requires specialized hardware for processing nor shows high computational complexity.

    \item We develop ScissionTL, which performs automated benchmarking for determining the optimal slicing point of the DNN model when the traffic is decreased by the TL. ScissionTL accurately finds the slicing point based on empirical data whereas other studies find an approximate solution with estimations.
    
\end{itemize}



    
    

The rest of this paper is organized as follows: Section~2 presents the background and related work. Section~3 presents the design of ScissionLite. Section~4 shows the performance evaluation results. Finally, we present our conclusions in Section~5.

\section{Background and Related Work}
\label{sec:background}
\begin{figure*}[t]
\centering
\includegraphics[width=0.9\textwidth]{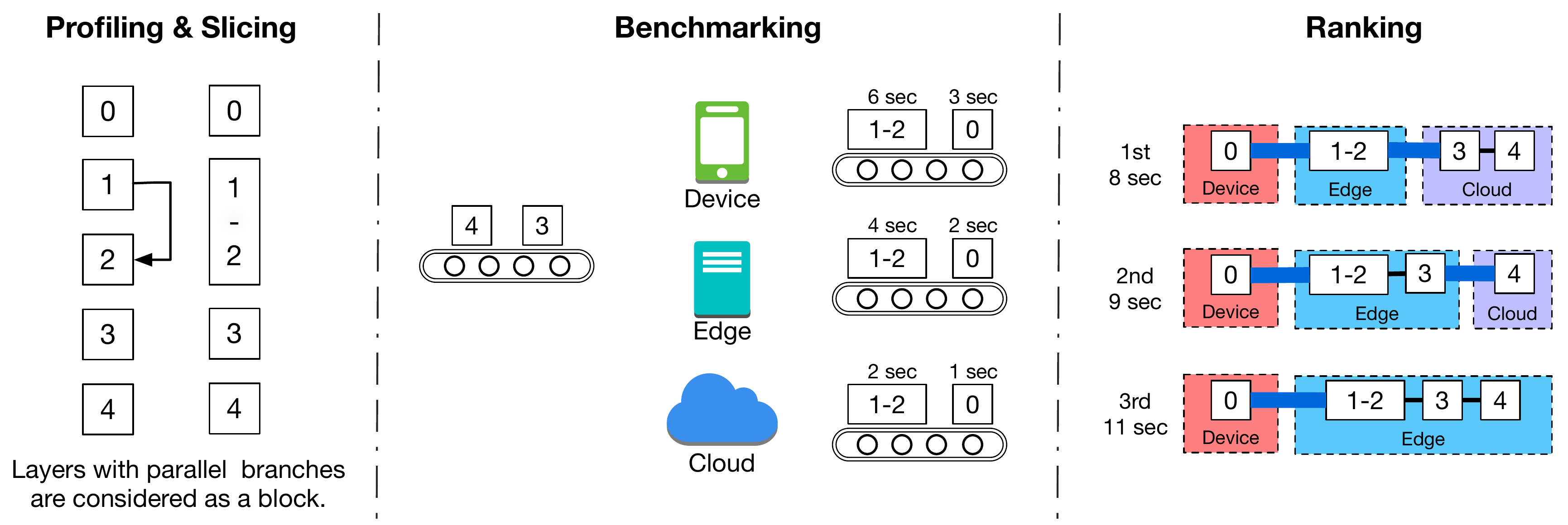}
\caption{Profiling, slicing, benchmarking and ranking in Scission}
\label{fig:scission}
\end{figure*}

In this section, we present the background research on which ScissionLite is developed and then present an overview of the related research. 

\subsection{Scission} 
\label{sec:Scission}

In ScissionLite, the transfer layer based benchmarking tool, namely ScissonTL, is based on Scission\footnote{Scission is available at \url{https://github.com/qub-blesson/Scission}}~\cite{lockhart2020scission} that determines the optimal slicing point for distributed deep learning inference.
The following design choices drove the development of Scission: (1) DNN slicing should be possible across different resource tiers in the device-edge-cloud continuum, (2) DNN slicing must identify optimally performing slices that can be distributed across resources, (3) DNN slicing must be rapid and based on empirical data rather than estimations because layers are extremely performance sensitive that cannot be easily predicted with estimations, and (4) DNN slicing should account for user-defined constraints, such as the target overall latency.

Scission is underpinned by a methodology involving profiling, slicing, benchmarking, and ranking as shown in Figure~\ref{fig:scission} and described below:

\textbf{Profiling:} The DNN is profiled to find suitable slicing points. All layers in the sequential DNN will be a slicing point. However, for a DNN model in which there are parallel branches, the parallel paths will need to be identified. The parallel branches are not sliced, but the layers within the branch are considered as a block. 

\textbf{Slicing:} This step ensures that the DNN is sliced into distinct sub-models with individual layers or blocks for the purposes of benchmarking.

\textbf{Benchmarking:} Each layer or block is benchmarked against hardware resources for obtaining the average execution time and the time taken to transfer data between the layers and blocks. 

\textbf{Ranking:} The DNN slices are ranked based on user-defined constraints, and a suitable DNN slice configuration can be chosen for deployment across the edge-cloud environment.

Although Scission supports different resource tiers including the cloud, the focus in this paper is on the device and edge tiers. Current processors offer powerful computation resources such as GPUs, and therefore they are leveraged at the edge without sending data to the cloud for better inference performance~\cite{wang2020convergence}.

\subsection{Related Work} 
\label{sec:Related Work}

Slicing and distributing DNNs across a combination of the device, edge, or cloud resources have performance benefits~\cite{edge-dnn-survey-1, wang2020convergence}. 
Slicing approaches rely on identifying a sequence of layers and mapping them onto resources so that the distributed DNN is optimized against the overall latency, ingress bandwidth, or a combination of these. 

Slicing requires the identification of the optimal slicing point, which in the literature is based on four different approaches. The first is an estimation-based approach in which the slicing point is determined based on an estimation of the performance of the layers on a target hardware platform~\cite{neurosurgeon, deepwear, musicalchair}. The second is based on integer linear programming, which is usually time consuming with the aim of minimizing inference latency and maximizing inference accuracy~\cite{jalad, jointdnn}. The third is based on a structural modification to the DNN. Although this approach is effective, they can be intrusive methods requiring modification of the underlying libraries~\cite{deepthings, modnn}. However, more fine-grain control over DNN slices can be obtained using this approach.
Finally, benchmarking based approaches for slicing DNNs are employed. These measurement-based approaches gather empirical data on the performance of each layer from target hardware resources. Scission, an underlying approach in ScissionTL presented in this paper, adopts a benchmarking-based approach~\cite{lockhart2020scission}. 

Given that ScissionLite uses the TL that is inserted in the DNN split point and is based on Scission, ScissionLite is a hybrid of structural modification-based and benchmarking-based DNN slicing approaches. 

Existing research presents methods to reduce the traffic amount in distributed deep learning inference. BottleNet~\cite{eshratifar2019bottlenet} compresses the output layer of a mobile device by using conventional compressors such as JPEG before sending the data to the cloud. The research was evaluated with ResNet50 and VGG19 in 3G, 4G, and WiFi environments that offer 1.1, 5.85, and 18.88 Mbps respectively for the upload bandwidth. Data compressors based on an autoencoder is presented~\cite{hu2020fast}. An autoencoder is an artificial neural network duplicating its inputs to the outputs with several convolutional layers and incurs additional computational complexity to the original DNN. This study utilized MobileNetV2 with CIFAR.

Compared to existing research, ScissionLite offers a lightweight compression technique using a down/upsampling network and a method to obtain the optimal slicing point in response to the network traffic change. In addition, it is evaluated by large-scale DNN models such as DenseNet201 and ResNet101 with the ImageNet dataset in a 5G environment.

\section{ScissionLite}
\label{sec:design}
\begin{figure}[t]\centering
\includegraphics[width=0.45\textwidth]{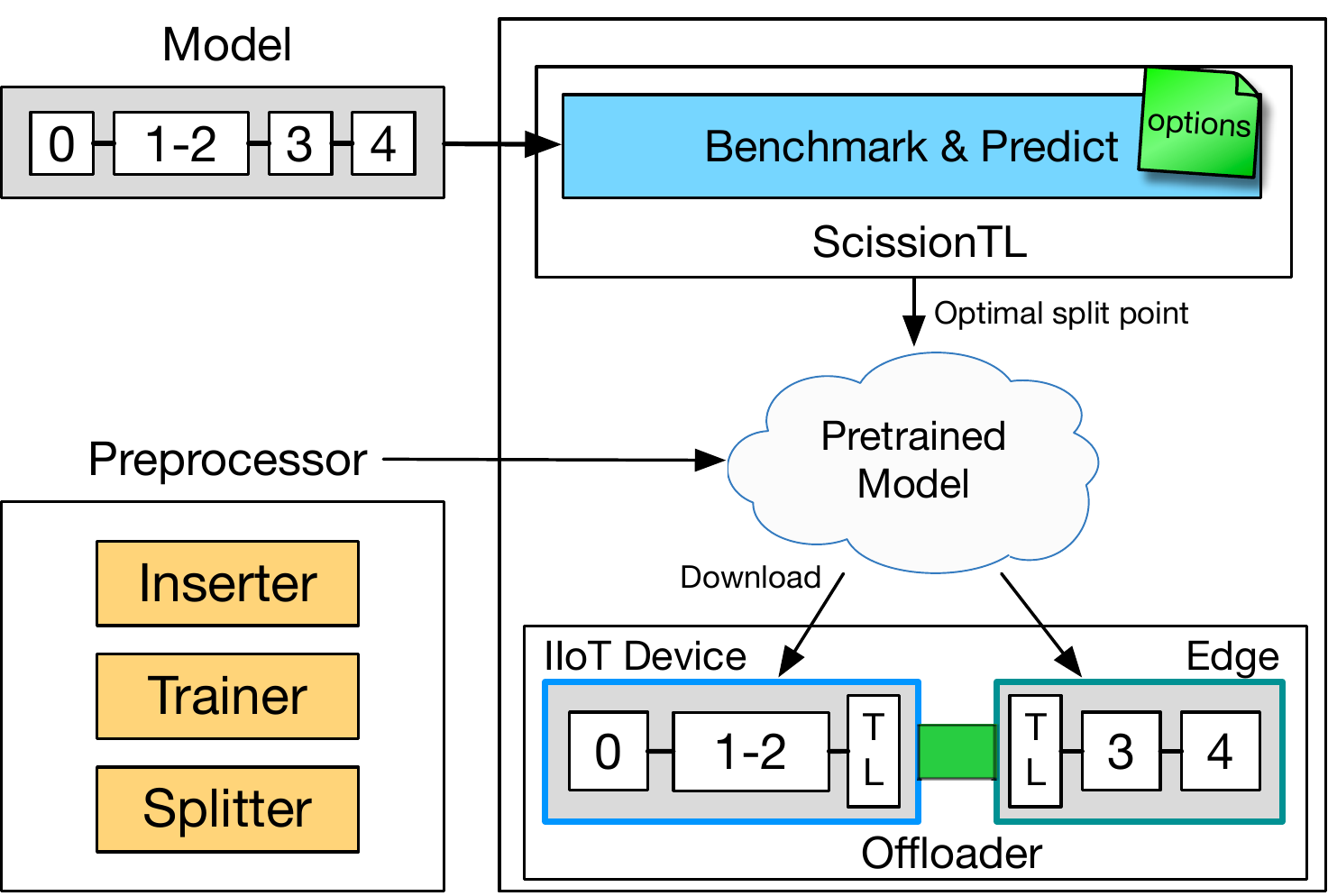}
\caption{Overall design of ScissionLite}
\label{fig:overall_design}
\end{figure}


In this section, we present the architecture of ScissionLite, which is a framework to accelerate edge-based distributed deep learning inference for IIoT.

\subsection{Overall Design}

ScissionLite inserts an additional traffic-aware layer called the Transfer Layer (TL) between the split point of the DNN. The TL decreases the amount of data transfer between the two entities by adopting a raw data encoding method. ScissionLite decides the optimal split point of the target DNN model considering the transfer amount reduced by the TL and injects the TL into the split point. For this purpose, ScissionLite develops three components including ScissionTL, the Preprocessor, and the Offloader as shown in Figure~\ref{fig:overall_design}. ScissionTL is a Scission-based tool for automated benchmarking that can decide the model's optimal slicing point when the TL is applied. The Preprocessor generates a new DNN model called the \emph{TLModel} that embeds the TL in the optimal slice point. The Offloader is a runtime platform that can deploy the TLModel and enable distributed inference utilizing both device and edge resources.

\subsection{Transfer Layer}
\label{sec:Transfer Layer}

The Transfer Layer (TL) is a small neural network layer that is embedded between the split point of the DNN for reducing the amount of data exchanged during communication. The TL is composed of the \emph{DeviceTL} and \emph{EdgeTL} layers respectively for the device and edge resources as shown in Figure~\ref{fig:overall_design}. The DeviceTL layer compresses the feature maps of the end layer of the sliced DNN on the IIoT device. The compressed data is then transferred through the cellular network connection. The EdgeTL layer expands the received data and passes the data to the starting layer of the remaining DNN on the edge. In this manner, the TL can decrease the amount of data transferred without significantly sacrificing inference accuracy.

The TL can be implemented by using an existing raw data encoding method such as a traditional image compression technique~\cite{liu2018deepn} (e.g., JPEG compression) or an autoencoder-based coding scheme~\cite{kurka2020deep}. These methods are a good candidate for realizing communication-efficient inference in edge-based machine learning systems. However, we identify that the former generally requires specialized hardware based on an application-specific integrated circuit (ASIC) to boost the compression and expansion process~\cite{ko2018edge}, and the latter demands significant training time for adapting the autoencoder and computation resources during inference due to the complexity of the autoencoder~\cite{kurka2020deep}. Since IIoT devices may have relatively performance-limited resources, a more efficient method is required for the TL.

To address this issue, we develop a new lightweight TL for IIoT using a down/upsampling neural network~\cite{giusti2013fast}, which can be performed on low power processors. With this thin TL, we can achieve acceptable training speed, inference time, and top-5 accuracy, which will be explained in Section \ref{sec:evaluation}. Although we present the design of the TL in the context of our own down/upsampling network, the TL is not specific to it and can be applied to other data encoding methods. Our implementations of the DeviceTL and EdgeTL layers are as follows:

\begin{figure}[t]\centering
\includegraphics[width=0.49\textwidth]{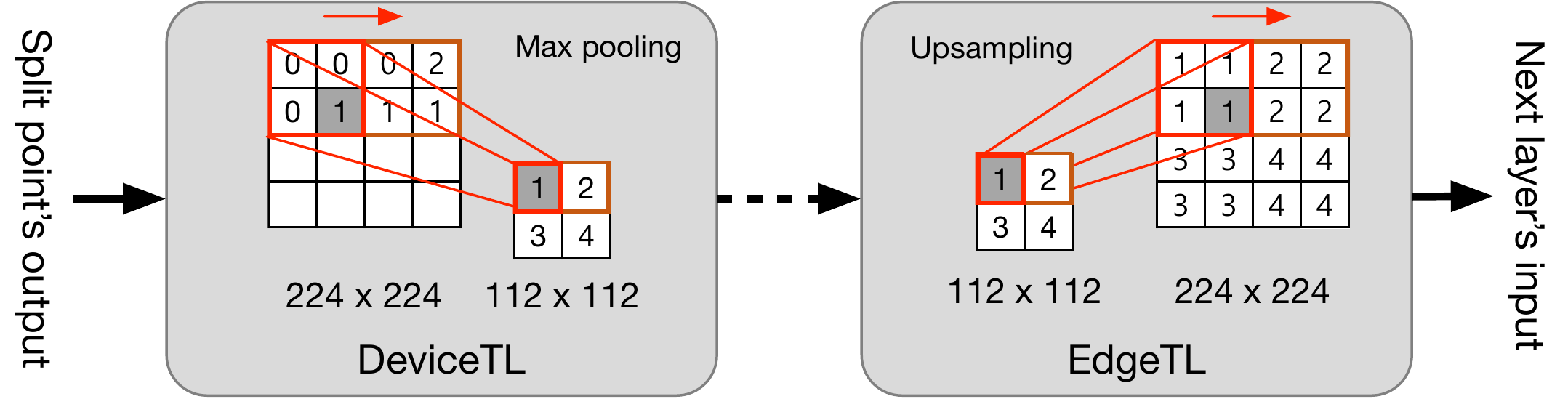}
\caption{Max pooling and upsampling of a neural network}
\label{fig:max_pooling}
\end{figure}

The DeviceTL layer compresses the output of the local IIoT device to a quarter by adopting a downsampling network. We implement the DeviceTL as a max pooling layer~\cite{giusti2013fast}, which sub-samples (or shrinks) each feature map of the last layer of the IIoT device. Each neuron in the DeviceTL is connected to a fewer number of neurons inside a small rectangular kernel in the last DNN layer as shown in Figure~\ref{fig:max_pooling}. We configure the kernel size as $2 \times $2 and select the maximum value in the kernel as an output value in the DeviceTL. Other neurons lower than the max value are dropped. We apply a stride of 2 with no padding to move the kernel back and forth across the feature map. The DeviceTL thus obtains sub-sampled feature maps whose size is a quarter of the last layer. A larger kernel size can improve the compression ratio but may incur a significant accuracy drop owing to information loss.

The EdgeTL layer expands the received data from the IIoT device by an upsampling neural network. The lost information caused by max pooling is replenished with a nearest-neighbor interpolation method, which selects the value from the nearest pixel and copies the selected value into the dropped neurons. This algorithm is computationally efficient. We have also considered other interpolation methods including bilinear and bicubic interpolation, but they resulted in an accuracy degradation than the nearest-neighbor interpolation.

\subsection{ScissionTL}
\label{ScissionTL}

As described in Section \ref{sec:Scission}, Scission automatically benchmarks DNN models on a target set of the device, edge, and cloud resources for deciding the optimal slicing point for maximizing inference performance. In this study, we develop \emph{ScissionTL}, which is a Scission-based tool for automated benchmarking that can acquire the optimal slicing point when the TL is used. This section presents a comparative analysis on Scission and ScissionTL and how ScissionTL finds the optimal slicing point.

The TL is an additional neural network layer that demands computation resources such as CPUs and GPUs during inference. The execution time for processing the TL is denoted by $E_{TL}$. As the TL is composed of the DeviceTL and the EdgeTL, $E_{TL}$ is calculated as follows:
\begin{equation}
\begin{aligned}
E_{TL} &= T(DeviceTL(Output_i)) \\ 
       &\quad+ T(EdgeTL(InputTL_i))
\end{aligned}
\end{equation}
where $T(x)$ denotes the execution time of running $x$. $Output_i$ is the output of the sliced DNN at splitting point $i$ in the device and becomes the parameter for the DeviceTL functionality. $InputTL_i$ is the downsampled result of $Output_i$ and is provided to the EdgeTL functionality.

As explained in Section \ref{sec:Transfer Layer}, we apply a lightweight max pooling and upsampling mechanism for the TL. Then, the computation cost, which is $E_{TL}$, is expected to be low. However, the computation burden would be high when a compute-intensive data encoding method such as an autoencoder is adopted.


Before network transmission, data or neurons compressed by the DeviceTL should be serialized to an appropriate data format such as Protocol Buffers (Protobuf) or JavaScript Object Notation (JSON). At the edge, the received data will be deserialized. This requires also computation resources. The execution time regarding (de)serialization, $S_{TL}$, is calculated as follows:
\begin{equation}
\begin{aligned}
S_{TL} &= T(Serial(OutputDown_i)) \\ 
       &\quad+ T(DeSerial(InputDownTL_i))
\end{aligned}
\end{equation}
where $Serial(OutputDown_i)$ serializes $OutputDown_i$, the down sampled result of $Output_i$, in the device, and $DeSerial(InputDownTL_i)$ deserializes $InputDownTL_i$, the serialized result of $OutputDown_i$, in the edge.

Original Scission's (de)serialization time, $S_{Orig}$, is calculated in the same way as follows:
\begin{equation}
\begin{aligned}
S_{Orig} &= T(Serial(Output_j)) \\ 
       &\quad+ T(DeSerial(InputOrig_j))
\end{aligned}
\end{equation}
where $Output_j$ is the output of the sliced DNN at splitting point $j$ in the device, and $InputOrig_j$ is the serialized result of $Output_j$.

The communication time between the device and the edge is denoted by $C_{TL}$, when the TL is applied. $C_{TL}$ is then calculated as follows:
\begin{equation}
C_{TL} = Latency + \frac{Size(OutputDown_i)}{Bandwidth} 
\end{equation}
where $Size(x)$ denotes the size of $x$, the network latency is $Latency$, and the network bandwidth is $Bandwidth$.

When the TL is not applied in original Scission, the communication time, $C_{Orig}$, is calculated as follows:
\begin{equation}
C_{Orig}  = Latency + \frac{Size(Output_j)}{Bandwidth} 
\end{equation}

Both communication times are affected by inherent network latency and the data transfer time, which can be obtained by dividing the size of the data by the network bandwidth. In the 5G cellular network, the upload bandwidth is as low as between 30 and 60 Mbps. Therefore, to decrease communication time, it is essential to reduce the amount of data transfer.

Consequently, $\Delta t$, which is the benefit of adopting the TL, is calculated as follows:
\begin{equation}
\Delta t = (S_{Orig} + C_{orig})-(E_{TL} + S_{TL} + C_{TL})
 \label{const4}
\end{equation}

The performance improvement that will be achieved is shown in Section~\ref{sec:evaluation}.

As presented in Figure \ref{fig:scission}, original Scission executes all possible split points individually for deciding the optimal slicing point. In addition to this, ScissionTL  embraces the above-mentioned parameters including $E_{TL}$, $S_{TL}$, and $C_{TL}$ for each slicing point. In the benchmarking and ranking phases, ScissionTL calculates the three parameters and uses these values to find the optimal slicing point.

\subsection{Preprocessor}
\label{Preprocessor}
The Preprocessor in ScissionLite prepares a pre-trained DNN model that employs the TL using the following three components: the Inserter, the Trainer, and the Splitter.

\textbf{Inserter:} The Inserter injects the TL in the split point of the DNN model. We have utilized pre-trained models from Keras Applications\footnote{\url{https://keras.io/api/applications/}}, which provide various DNN models with pre-trained weights. The Inserter loads the target model and divides the model into two for putting the TL between them. Afterward, it joins the sub-models together for retraining. The new model created by the Inserter is called the \emph{TLModel}.

\textbf{Trainer:} The Trainer retrains the TLModel for preventing an accuracy drop. We use a pre-trained model from Keras Applications, but the pre-trained weights are not aware of the TL, which can lead to a significant accuracy drop. Existing neurons then need to learn how to cooperate with new neurons in the TL. As the TL is lightweight, existing layers can be adapted to the TL fast during retraining. The top-5 accuracy and retraining time of the TLModel will be presented in Section~\ref{Inference Accuracy}.

\textbf{Splitter:} The Splitter divides the TLModel into two for the IIoT device and the edge. For the device-side model, the Splitter piles up the layers until the $DeviceTL$ layer and exports them as a single Keras model. Similarly, the edge-side model starts from the $EdgeTL$ layer and ends at the last fully-connected layer.

\subsection{Offloader}
\label{Offloader}

The Offloader in ScissionLite is an inference platform that deploys the TLModel on the device and the edge and establishes the connection between them. It consists of the Runtime and the Communicator.

\textbf{Runtime:} In the IIoT device, we use the TensorFlow runtime to execute the sliced TLModel. On the edge side, we utilize the NVIDIA Triton inference server\footnote{\url{https://developer.nvidia.com/nvidia-triton-inference-server}}, which is open-source serving software and optimized for edge deployment, for executing the remaining TLModel. The output data of $DeviceTL$ is transferred to the $EdgeTL$ layer in Triton through the Communicator. After the inference task is completed in the edge, the result is returned to the IIoT device again using the Communicator.

\textbf{Communicator:} The Communicator connects each Runtime in the IIoT device and the edge. Both REST and gRPC protocols can be utilized as a communication method, but we adopted gRPC because it is based on HTTP/2 and is faster than REST with HTTP/1.1. gRPC uses Protocol Buffers (Protobuf) as the message interchange format. The Communicator implements the Device and Edge converters in order to serialize the output of $DeviceTL$ to Protobuf and deserialize the Protobuf to the input data of $EdgeTL$.

\section{Evaluation}
\label{sec:evaluation}
We present the performance evaluation results of ScissionLite in this section. 

\subsection{Experimental Environment}

\begin{table}[t]\centering
\caption{Test bed configurations}
\includegraphics[width=0.45\textwidth]{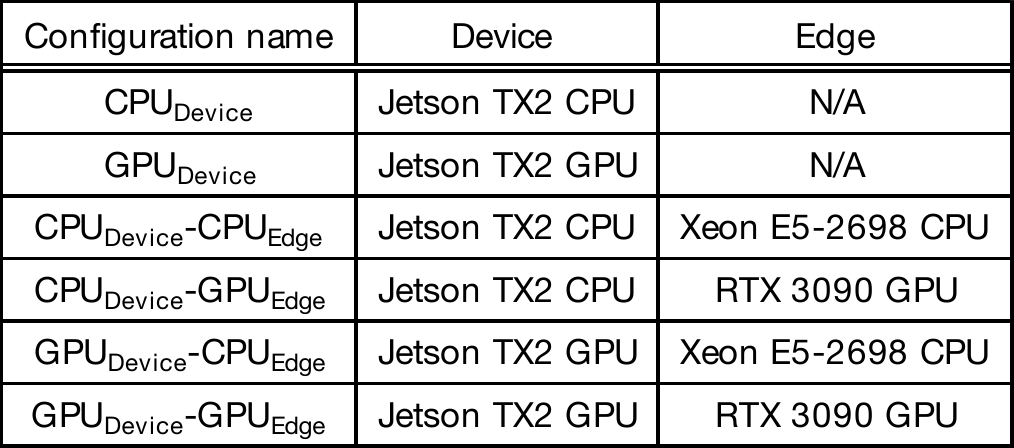}
\label{table:Test bed configurations}
\end{table}

\begin{figure}[t]\centering
\includegraphics[width=0.45\textwidth]{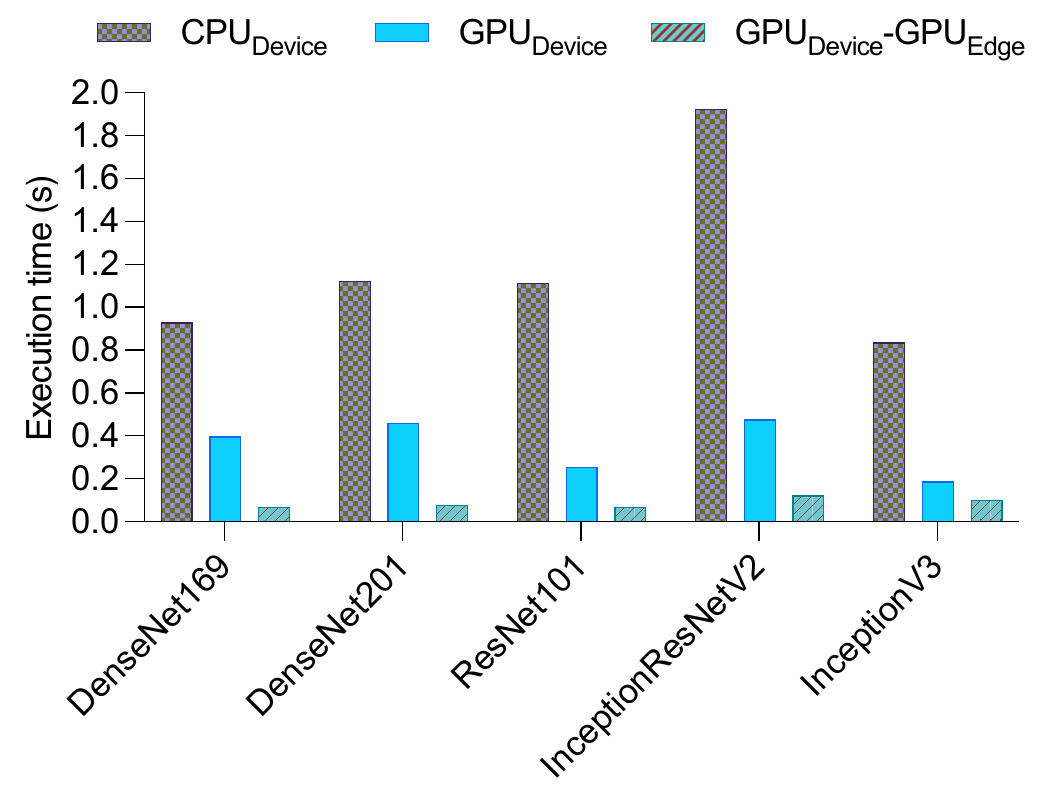}
\caption{Best performance of various DNN models obtained from the CPU\textsubscript{Device}, GPU\textsubscript{Device}, and GPU\textsubscript{Device}-GPU\textsubscript{Edge} configurations}
\label{fig:feasibility}
\end{figure}

\begin{figure*} [!t]
\centering
\includegraphics[width=0.90\textwidth]{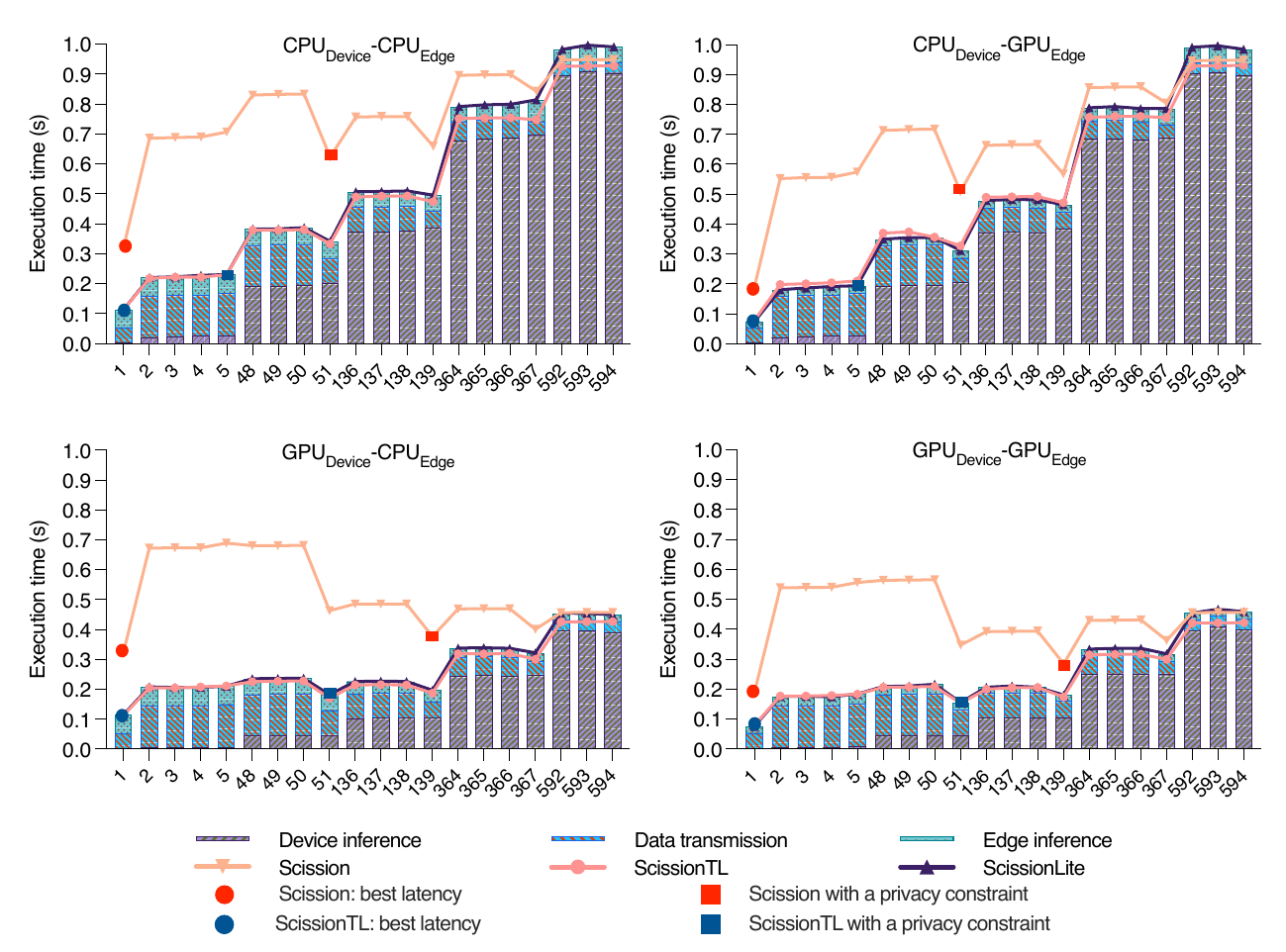}
\caption{Slice-by-slice analysis of DenseNet169 in Scission, ScissionTL, and ScissionLite with four test bed configurations.}
\label{fig:densenet}
\end{figure*}

\begin{figure*} [!t]
\centering
\includegraphics[width=0.90\textwidth]{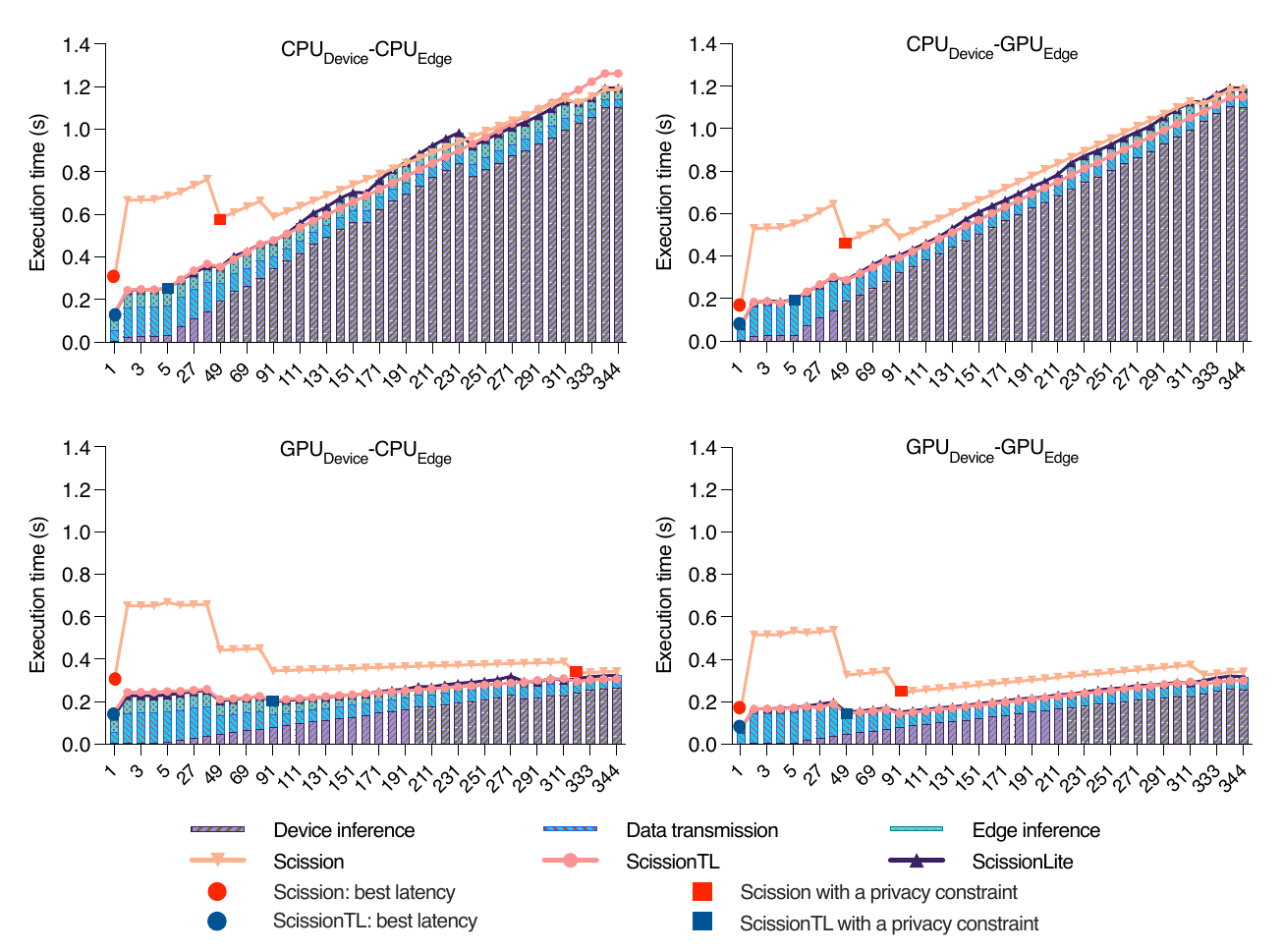}
\caption{Slice-by-slice analysis of ResNet101 in Scission, ScissionTL, and ScissionLite with four test bed configurations.}
\label{fig:resnet}
\end{figure*}

We implemented ScissionLite with an IIoT device and an edge server for distributed inference. As the IIoT device, we used an NVIDIA Jetson TX2, which is an embedded AI computing device that equips two dual-core and quad-core ARM CPUs and an NVIDIA Pascal-family GPU having 8 GB of memory. The edge server is an Intel Xeon E5-2698 CPU platform with twenty 2.2 GHz cores and an NVIDIA GeForce RTX 3090 GPU having 24 GB of memory. The IIoT device is directly connected to the edge server via 1 GbE link. We adjusted the upload bandwidth and latency of the network connection as 30 -- 60 Mbps and 30 ms respectively by exploiting the Linux Traffic Control~\cite{hubert2002linux} tool. These values emulate the commercial 5G network environment based on the latest research~\cite{narayanan2020first}. 

We employed the TensorFlow runtime version 2.3 in the IIoT device and NVIDIA Triton as an inference server in the edge. We obtained deep learning models with pre-trained weights from Keras Applications\footnote{\url{https://keras.io/api/applications/}}. We used five pre-trained production DNNs, namely DenseNet169, DenseNet201, ResNet101, InceptionResNetV2, and InceptionV3 for evaluation. We also employed ImageNet~\cite{deng2009imagenet} as an image database. The test bed configurations reflecting an operational IIoT environment are presented in Table~\ref{table:Test bed configurations}. We combined possible hardware resource types in the IIoT device and the edge server.



\subsection{Inference Latency}

\begin{figure*} [!t]
\centering
\includegraphics[width=0.90\textwidth]{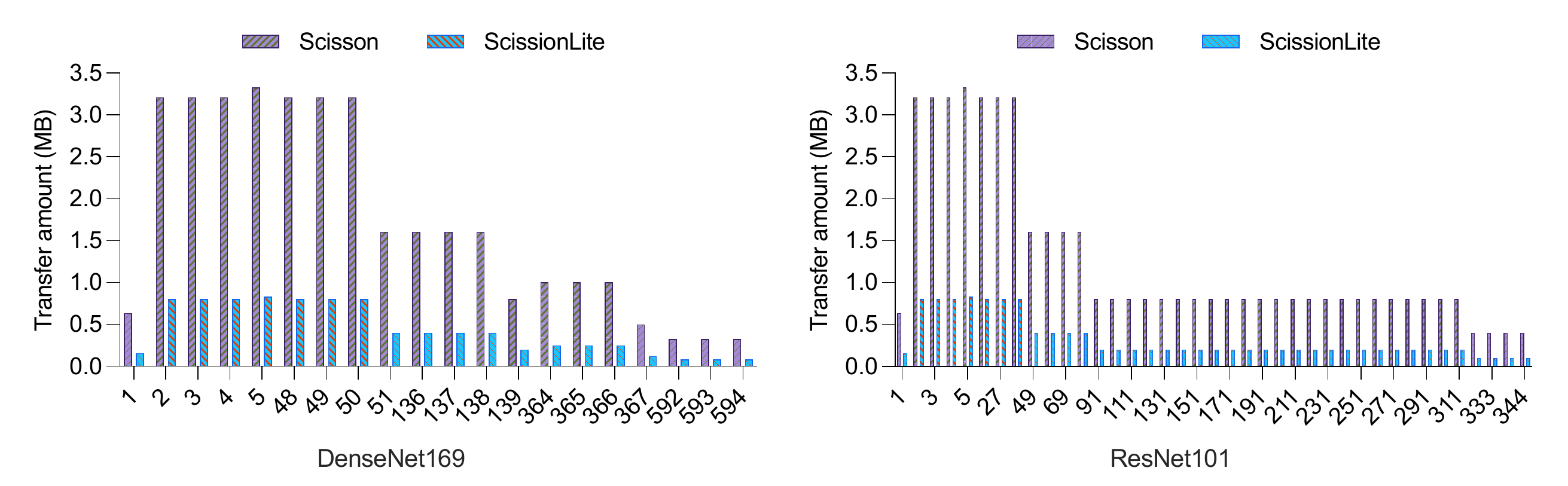}
\caption{Transferred data amount of DenseNet169 and ResNet101 at each split point in Scission and ScissionLite.}
\label{fig:data_amount}
\end{figure*}

\begin{figure}[t]\centering
\includegraphics[width=0.45\textwidth]{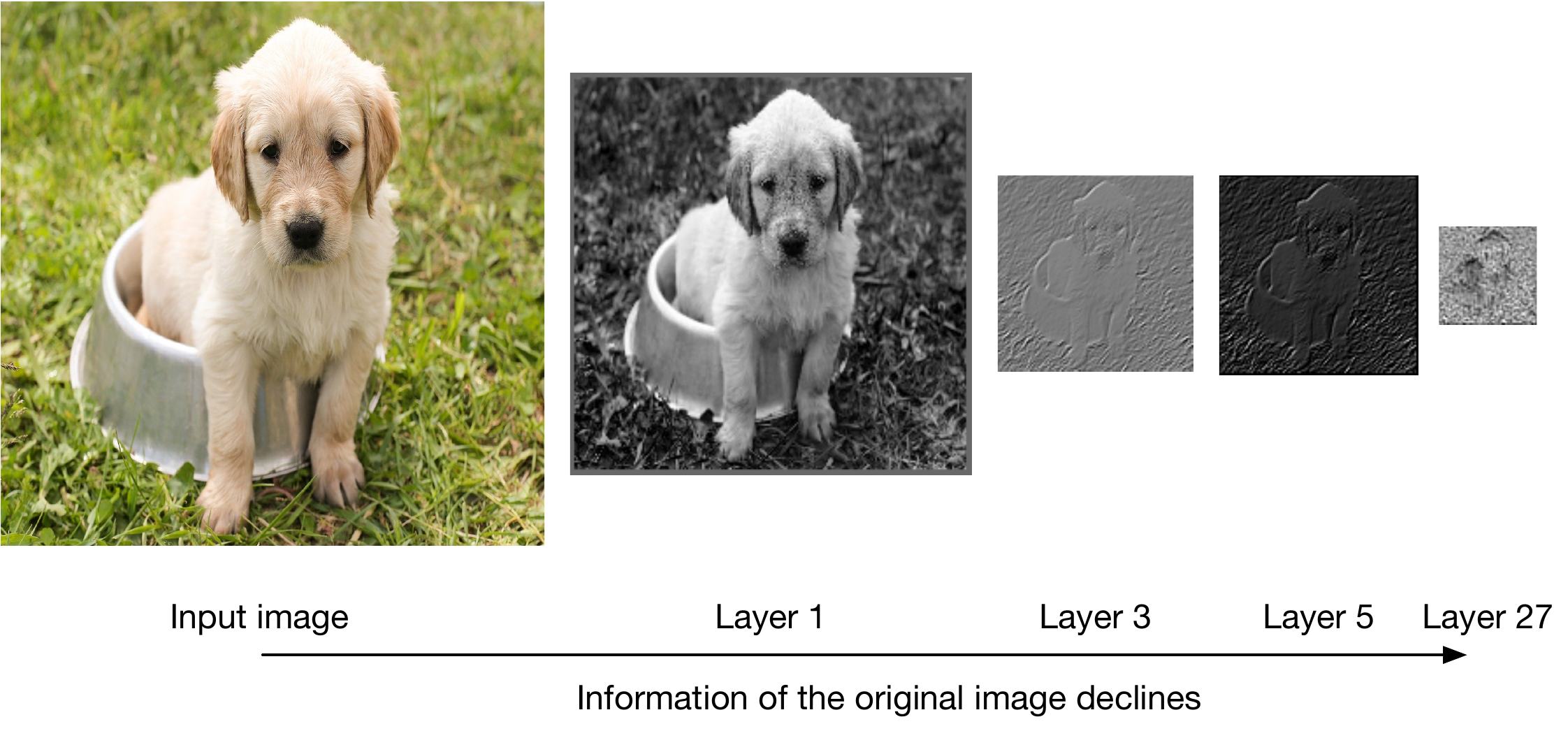}
\caption{Intermediate representations in front layers of ResNet101}
\label{figure:resnet_layer_image}
\end{figure}

\begin{figure*} [!t]
\centering
\includegraphics[width=0.90\textwidth]{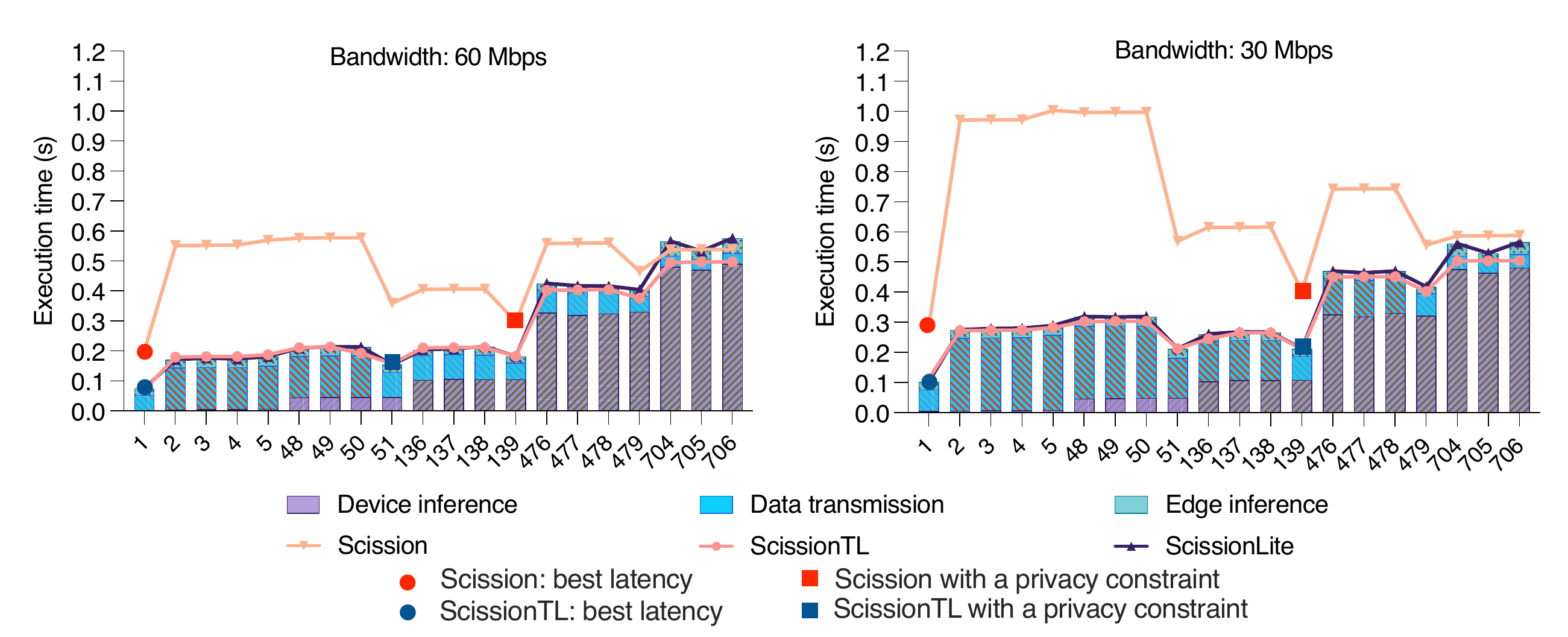}
\caption{Inference latency of DenseNet201 when the network bandwidth is given as 60 and 30 Mbps respectively.}
\label{figure:bandwidth}
\end{figure*}

ScissionLite accelerates distributed deep learning inference by inserting the TL between the split point. ScissionLite first obtains the optimal split point from ScissionTL for minimizing inference latency and runs the sliced DNN model with the Offloader. 

\textbf{Feasibility of distributed inference:} Figure~\ref{fig:feasibility} shows the best inference latency of the five DNN models respectively executed in the CPU\textsubscript{Device}, GPU\textsubscript{Device}, and GPU\textsubscript{Device}-GPU\textsubscript{Edge} configurations in Table~\ref{table:Test bed configurations}. The first two configurations do not employ DNN model slicing so that the DNN models were run in the local IIoT device using its embedded CPU and GPU. The last configuration utilizes DNN model slicing with ScissionLite. As shown in the figure, ScissionLite improves the performance of the DNN models up to 16 and 5.7 times compared to the CPU\textsubscript{Device} and GPU\textsubscript{Device} configurations respectively. This result shows that DNN model slicing is effective in edge-based IIoT.

\textbf{Transfer Layer (TL) computation overhead:} ScissionLite adopts the TL to decrease the volume transferred to a quarter of the size. The TL compresses the output of the IIoT device by using a downsampling network and expands the compressed data with an upsampling network at the edge. The scheme is computationally lightweight for an IIoT device. The DeviceTL takes at most 300 $\mu$s on the device GPU and 2,500 $\mu$s on the device CPU. The EdgeTL takes at most 200 $\mu$s on the edge server GPU. Therefore, the overhead of adopting the TL is low.

\textbf{Slice-by-slice analysis:} In this experiment, we provide a slice-by-slice analysis of DenseNet169 in Figure~\ref{fig:densenet} and ResNet101 in Figure~\ref{fig:resnet} in Scission, ScissionTL, and ScissionLite with four test bed configurations. We represent the Offloader in ScissionLite, which is an actual runtime platform, as ScissionLite in the following explanations. The upload bandwidth and network latency are configured as 57 Mbps and 28 ms respectively, which are the peak performance of commercial 5G~\cite{narayanan2020first}. 

Each sub-figure presents the total end-to-end latency when the split point or layer changes from the first to the last layer; less important split points are not shown in the figure. In the case of ScissionLite, the profiling result in each split point is shown as a stacked column chart. The optimal split points decided by Scission are denoted by a red circle and square. ScissionTL's decisions are shown by the blue symbols.

As shown in Figure~\ref{fig:densenet} and ~\ref{fig:resnet}, the latency values of ScissionTL and ScissionLite are converged across the split points in each test bed configuration. This means that both ScissionTL and ScissionLite are synchronized. Therefore, the optimal point suggested by ScissionTL without testing all split points in the production platform is trustworthy. This is useful when the network environment or a user-defined constraint is changed in the IIoT system. For example, when the network bandwidth of 5G is significantly increased, ScissionTL can decide on a new slicing point rapidly with existing benchmarked results. In this case, the new DNN can be deployed to the IIoT production platform without further testing.

Based on the optimal points suggested by Scission and ScissionTL in DenseNet169 and ResNet101, ScissionLite performs faster than Scission, because ScissionLite adopts the traffic-aware layer between the split point. ScissionLite can improve the performance of distributed inference by up to 2.8, 2.6, 2.8, and 2.5 times in DenseNet169 respectively with the CPU\textsubscript{Device}-CPU\textsubscript{Edge}, CPU\textsubscript{Device}-GPU\textsubscript{Edge}, GPU\textsubscript{Device}-CPU\textsubscript{Edge}, and GPU\textsubscript{Device}-GPU\textsubscript{Edge} configurations, and 2.31, 2.42, 2.24, and 2.25 times in ResNet101 with the same configurations, compared to Scission.

In all configurations, both Scission and ScissionTL suggest Layer 1 as the optimal point in DenseNet169 and ResNet101. This is because the amount of data transfer at Layer 1 is the smallest among early and middle layers as shown in Figure~\ref{fig:data_amount}. Scission and ScissionTL then decide to process most of DNN layers at the edge server to exploit more powerful CPUs and GPUs. 

Transferring the output of Layer 1 can benefit from privacy preservation than sending the original image to the edge~\cite{lockhart2020scission}. However, when the output of a later layer is sent to the edge, security and privacy can be enhanced. As the split layer is higher, the intermediate representations contain less information on the original image as shown in Figure~\ref{figure:resnet_layer_image}.

To enhance privacy, we impose a user constraint by configuring the split point as five or more; this value can be changed according to the DNN. In most cases, Scission suggests a later layer as the optimal point (red square symbol) compared with ScissionTL (blue square) in DenseNet169 and ResNet101. Scission considers the significantly increased data amount right after Layer 1 (Figure~\ref{fig:data_amount}), and investigates later points that can meet both low traffic overhead and computation offloading benefit. However, ScissionTL is less affected by the data traffic amount due to the TL, and therefore it suggests an earlier layer than Scission. For example, ScissionTL recommends Layer 51, whereas Scission proposes Layer 139 in the GPU\textsubscript{Device}-GPU\textsubscript{Edge} configuration in DenseNet169.

\textbf{Different network bandwidth:} Figure~\ref{figure:bandwidth} shows the inference latency of DenseNet201 with the GPU\textsubscript{Device}-GPU\textsubscript{Edge} configuration when the network bandwidth is given as 60 and 30 Mbps respectively. Scission shows increased latency between Layer 2 and 50 at 30 Mbps owing to large data transfer (Figure~\ref{fig:data_amount}). However, ScissionLite shows stable performance even with limited bandwidth owing to the TL.

\subsection{Inference Accuracy}
\label{Inference Accuracy}

\begin{table}[t]\centering
\caption{Top-5 accuracy of the original Keras model and the retrained TLModel in ScissionLite}
\includegraphics[width=0.45\textwidth]{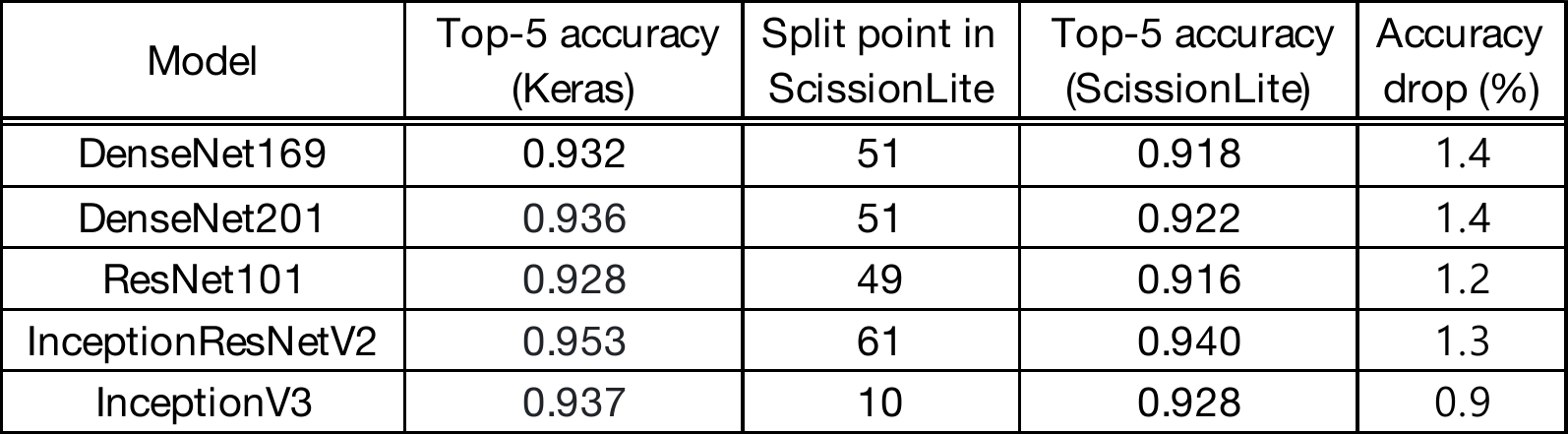}
\label{table:accuracy}
\end{table}

In ScissionLite, the TL implements a down/upsampling neural network, which is computationally efficient for IIoT devices. However, as the feature maps of the last DNN layer are subsampled in the DeviceTL, the TL causes some information loss. In order to prevent an accuracy drop, the Trainer retrains the split model embedding the TL (TLModel), as explained in Section~\ref{Preprocessor}. The Trainer employs the stochastic gradient descent (SGD) optimizer with 0.001 of a learning rate for retraining. The TLModel is adapted to the new neurons in the TL during retraining and converges to minimum loss rapidly. This takes about between 5 and 10 hours with a single NVIDIA GeForce RTX 3090 GPU. The training speed can be boosted by using multiple GPUs in a cloud data center. 

Table~\ref{table:accuracy} compares the top-5 accuracy of the original Keras model and the retrained TLModel in ScissionLite. Both accuracy values were measured with the ImageNet validation dataset\footnote{\url{https://keras.io/api/applications/}}. The result shows that the retrained TLModel provides acceptable inference accuracy with just 0.9 -- 1.4\% of a drop. This highlights that the TL relieves the communication overhead without deteriorating the overall accuracy significantly.

\section{Conclusions}
\label{sec:conclusions}
Edge computing offers the distinct advantage of bringing compute resources closer to devices and sensors where data is generated for reducing the overall application latency. The Industrial Internet of Things (IIoT) is one area that benefits from using the edge. Given that IIoT devices may have relative resource limitations when compared to the edge, workloads can be offloaded from the device to the edge to accelerate data processing. For example, Deep Neural Network (DNN) applications used in IIoT can be sliced and distributed across the IIoT device and the edge for reducing the total inference time. 

One of the key challenges when slicing DNNs and distributing them is finding the optimal slicing point while accounting for the inter-layer data transfer sizes. An optimal slicing point may require a large amount of data to be transferred between the DNN slices on the device and the edge. This is problematic when faced with variable network conditions between the device and the edge. 

This paper presents an end-to-end traffic-aware framework, namely ScissionLite that finds the optimal DNN slicing point but at the same time reduces the amount of data transferred between the slices across the device and the edge. A novel method that uses a lightweight transfer layer using a down/upsampling network for performance-limited IIoT devices is developed. 

ScissionLite is tested on five major production DNNs in a lab-based physical test bed comprising GPUs. The experimental results obtained highlight that DNN inference can improve performance by up to 16 times compared with the local execution while only incurring a 0.9\% -- 1.4\% accuracy drop. The use of the transfer layer is shown to have a negligible overhead. ScissionLite demonstrates the feasibility of the transfer layer approach for accelerating DNN inference for IIoT applications. 

\ifCLASSOPTIONcompsoc
  \section*{Acknowledgments}
\else
  \section*{Acknowledgment}
\fi
The authors are grateful to the anonymous reviewers for their valuable comments and suggestions. This work was supported by the National Research Foundation of Korea (NRF) grant funded by the Korean government (MSIP) (No. NRF-2019R1C1C1011068). The last author was supported by a Royal Society Short Industry Fellowship. 

\ifCLASSOPTIONcaptionsoff
  \newpage
\fi

\bibliographystyle{IEEEtran}
\bibliography{IEEEabrv,paper-v1}

\begin{IEEEbiography}[{\includegraphics[width=1in,height=1.25in,clip,keepaspectratio]{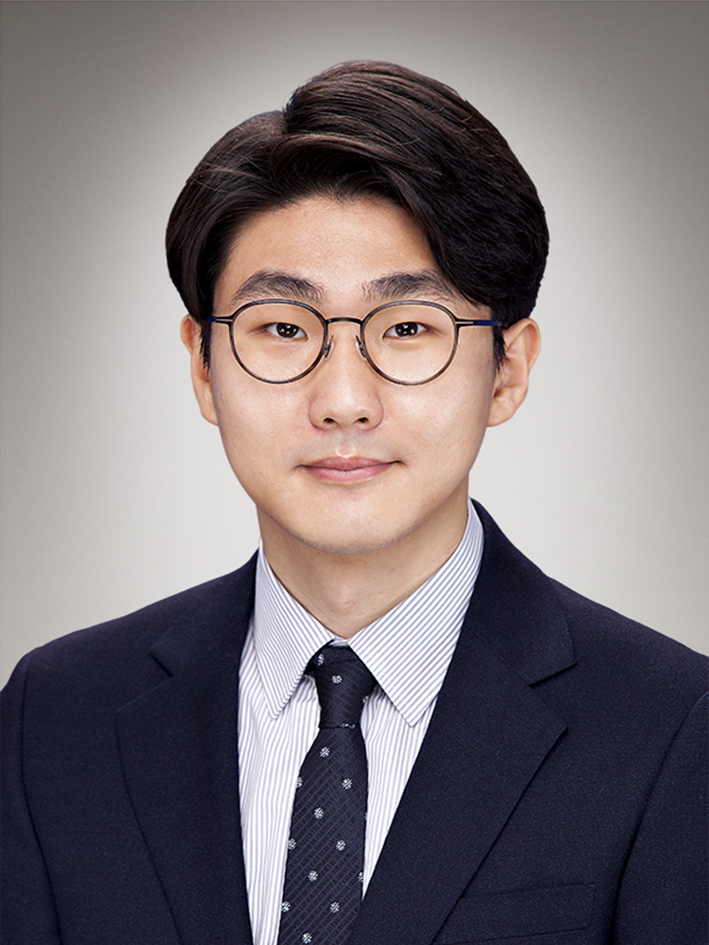}}]{Hyunho Ahn}
received the B.E. degree in Electrical and Electronics Engineering from Chung-Ang University, Seoul. He is now a master student in the School of Electrical and Electronics Engineering, Chung-Ang University, Seoul, Korea. His current research interests include edge computing and deep learning.
\end{IEEEbiography}

\begin{IEEEbiography}[{\includegraphics[width=1in,height=1.25in,clip,keepaspectratio]{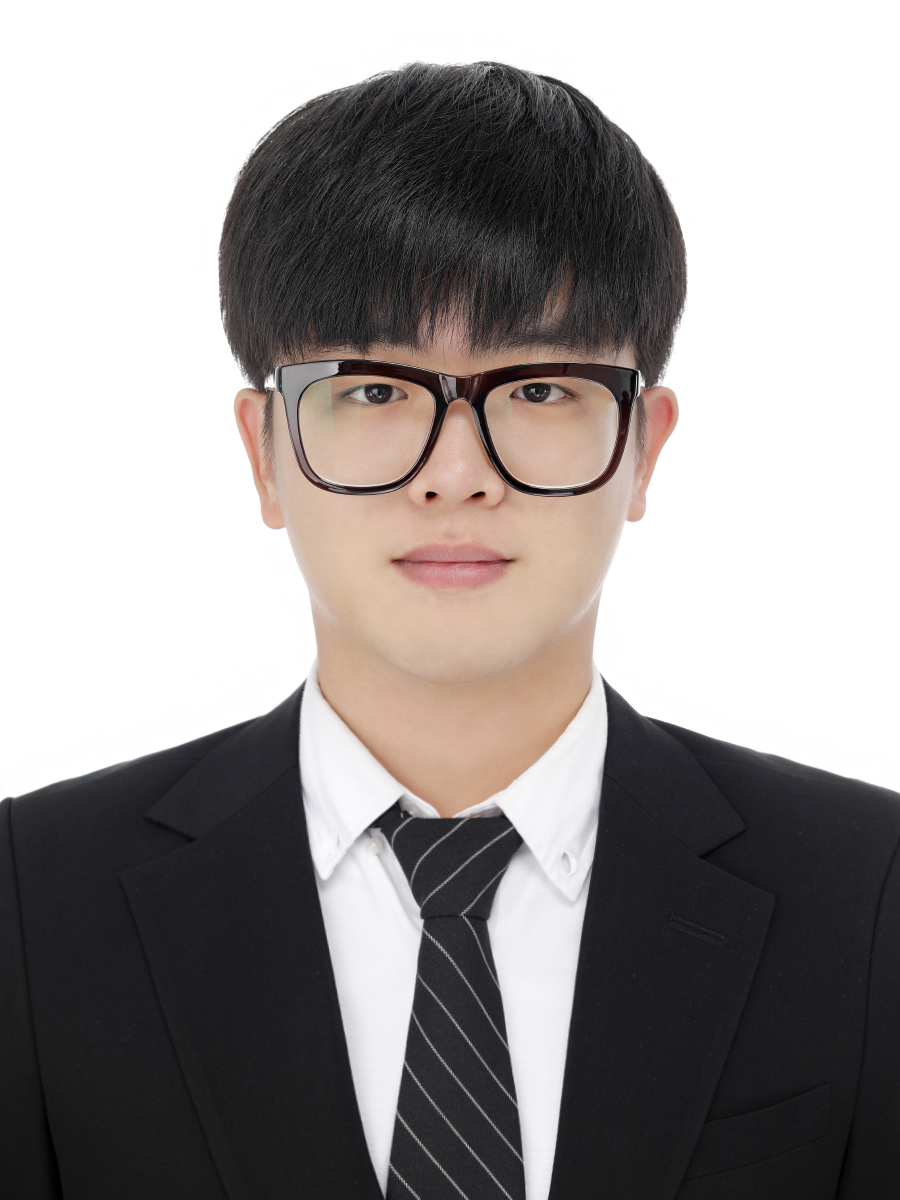}}]{Munkyu Lee}
received the B.E. and M.E. degrees in Electrical and Electronics Engineering from Chung-Ang University, Seoul. He is now a Ph.D. candidate in the School of Electrical and Electronics Engineering, Chung-Ang University, Seoul, Korea. His current research interests include system virtualization, deep learning, and GPU architectures.
\end{IEEEbiography}

\begin{IEEEbiography}[{\includegraphics[width=1in,height=1.25in,clip,keepaspectratio]{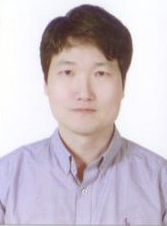}}]{Cheol-Ho Hong}
received the BS, MS, and PhD degrees in Computer Science from Korea University, Seoul. He worked as a Research Professor in Korea University from 2013 to 2015 and as a Research Fellow in Queen's University Belfast from 2015 to 2018. He is now an Associate Professor in the School of Electrical and Electronics Engineering, Chung-Ang University, Seoul, Korea. His current research interests include system virtualization, edge computing, deep learning, and GPU architectures.
\end{IEEEbiography}

\begin{IEEEbiography}[{\includegraphics[width=1in,height=1.25in,clip,keepaspectratio]{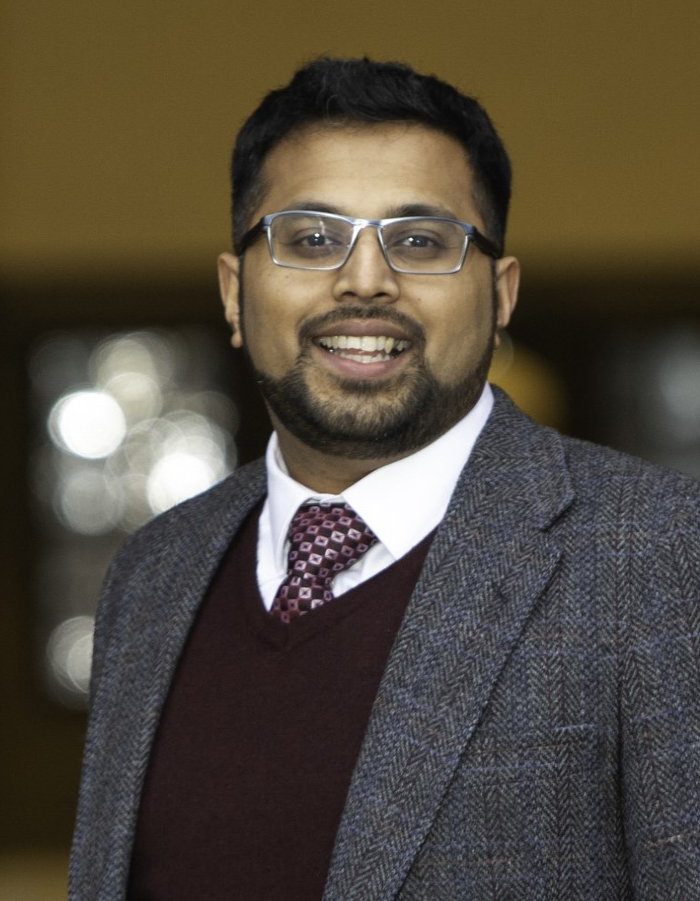}}]{Blesson Varghese}
received the PhD degree in Computer Science from the University of Reading, UK. He worked as a researcher at the Dalhousie University, Canada, and the University of St Andrews previously. He is now a Senior Lecturer (Associate Professor) in the School of Electronics, Electrical Engineering and Computer Science, Queen's University Belfast, UK and a Royal Society Short Industry Fellow. His current research interests include developing and analyzing state-of-the-art cloud-edge systems and distributed machine learning. More information at \url{www.blessonv.com}.
\end{IEEEbiography}




\end{document}